\documentclass{article}
\usepackage{spconf,amsmath,graphicx,amssymb}
\usepackage{hyperref}

\title{Automating detection of papilledema in pediatric fundus images \\ with explainable machine learning}

\name{Kleanthis Avramidis$^1$,\; Mohammad Rostami$^1$,\; Melinda Chang$^2$,\; Shrikanth Narayanan$^1$}
\address{ $^1\;$Signal Analysis and Interpretation Lab, University of Southern California, Los Angeles, CA 90089 \\ $^2\;$USC Roski Eye Institute, University of Southern California, Los Angeles, CA 90089}

\begin{document}
\ninept
\maketitle
\begin{abstract} % 100 to 150 words
Papilledema is an ophthalmic  neurologic disorder in which increased intracranial pressure leads to swelling of the optic nerves. Undiagnosed papilledema in children may lead to blindness and may be a sign of life-threatening conditions, such as brain tumors. Robust and accurate clinical diagnosis of this syndrome can be facilitated by automated analysis of fundus images using deep learning, especially in the presence of challenges posed by pseudopapilledema that has similar fundus appearance but distinct clinical implications. We present a deep learning-based algorithm for the automatic detection of pediatric papilledema. Our approach is based on optic disc localization and detection of explainable papilledema indicators through data augmentation. Experiments  on real-world clinical data demonstrate that our proposed method is effective with a diagnostic accuracy comparable to expert ophthalmologists\footnote{Code available at \href{https://github.com/klean2050/papilledema}{https://github.com/klean2050/papilledema}}.

\end{abstract}
\begin{keywords}
human-centered AI, model explainability, papilledema, pseudopapilledema,  multi-view learning\vspace{-0.1cm}
\end{keywords} 

\section{Introduction}
\label{sec:intro}

\begin{figure}
    \centering
    \includegraphics[scale=0.28]{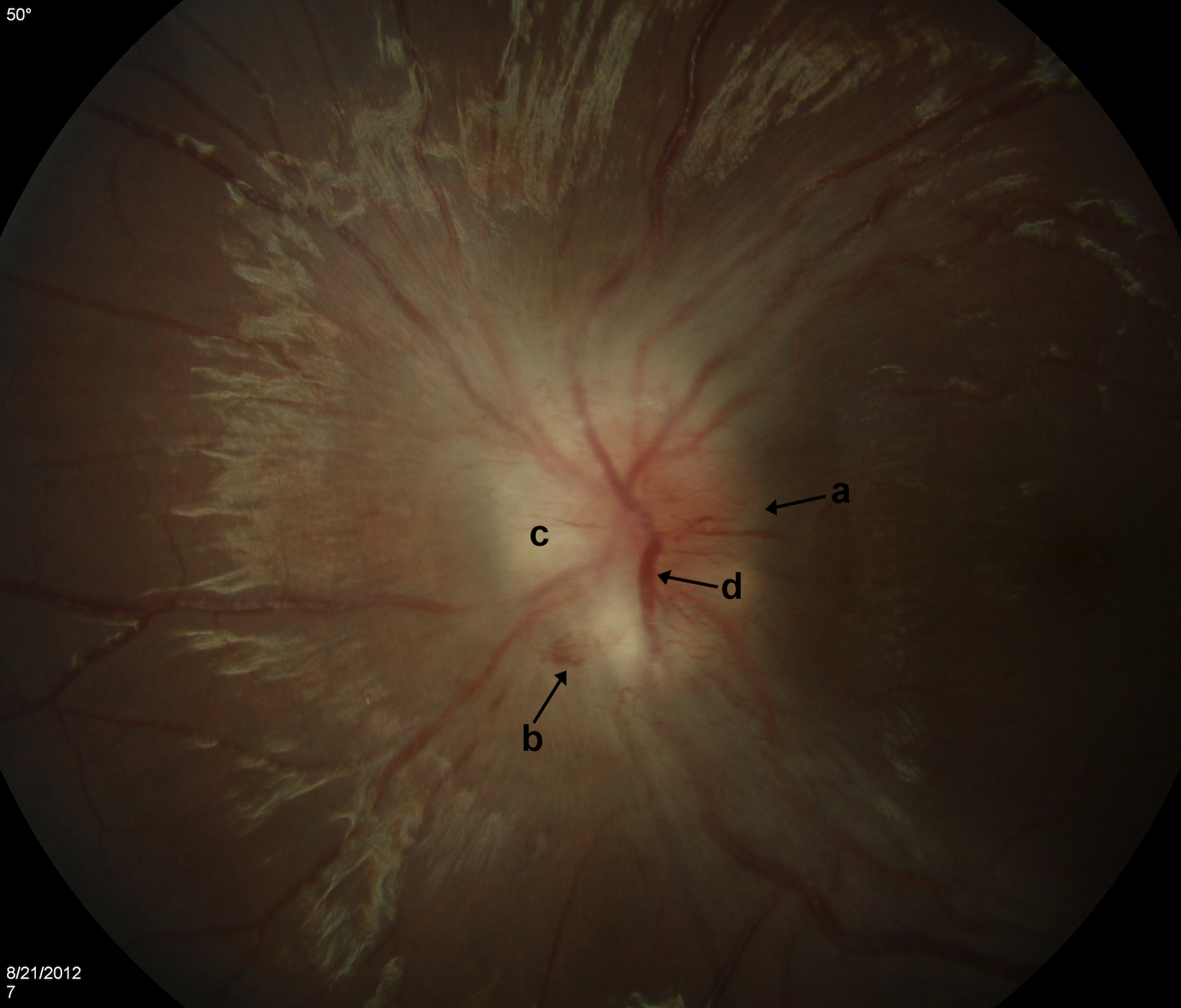}
    \vspace{-0.2cm}
    \caption{Papilledema clinical indicators in a fundus photo of the left eye of a patient with papilledema, demonstrating a) swelling of the retinal nerve fiber layer, b) peripapillary hemorrhage, c) elevation, and d) congestion of the retinal venule on the optic nerve.}
    \label{fig:features}
    \vspace{-0.2cm}
\end{figure}

\vspace{-0.1cm}The human visual system develops over several years during early childhood and fully matures around age seven years. During this critical period of  development, there are potential clinically-relevant issues that may emerge or manifest. One such issue, papilledema, occurs when increased intracranial pressure leads to swelling, or edema, of the optic nerves. It is crucial to diagnose pediatric papilledema early, because it could indicate a serious underlying neurologic disorder, such as a brain tumor.
%It is also important to diagnose papilledema promptly because
Untreated papilledema may also lead to vision loss or blindness. Early diagnosis of papilledema using fundus imaging enables clinicians to perform prognosis and timely therapeutic intervention. Fundus imaging uses a specialized low-power microscope camera to photograph the interior surface of the eye, including the retina and the optic disc (Fig.~\ref{fig:features}). A major challenge that is faced in the robust and accurate detection of pediatric papilledema is
%that potential confounds due to
the presence of {\em pseudopapilledema}
%appear more similar to papilledema in children. Pseudopapilledema
\cite{pseudo}, an abnormality that occurs when the optic disc is elevated without swelling of the optic nerve. As a result, no threatening disorders are associated with pseudopapilledema. However, due to similar clinical presentations, yet significantly different treatment protocols,  distinguishing between pseudopapilledema and papilledema is critical for efficient care. This task
%, however, is deemed to be an especially challenging task
is especially challenging using conventional diagnostic tools like fundus images, even for trained ophthalmologists~\cite{ahn2019accuracy}.

The availability of large clinical datasets along with high-performance processors and advances in machine learning systems enable us to design image processing algorithms that can automate the clinical diagnosis procedures~\cite{milea2020artificial,hernandez2021improving,degadwala2021eye}. However, a major drawback% -- and at the same time, a drawback -- 
of deep learning is that the feature extraction process from input images can be automated using an end-to-end blind supervised training procedure. As a result, interpreting the decision process by such a deep neural network (DNN) is challenging. Moreover, empirical explorations have demonstrated that, despite high-accuracy performances, DNNs do not necessarily use a human-centered and interpretable decision process. In clinical decision making, which is a high stakes setting, it is thus often difficult for experts to rely on ``black-box" DNN-based methods. To address this shortcoming, in this paper we develop a multi-branch deep neural network to classify an input fundus image as a papilledema or a pseudopapilledema case. Our proposed model extracts interpretable features from the input through data augmentation and multi-view training, thus leading to a decision process similar to that of expert ophthalmologists. \vspace{-0.4cm}

\section{Background}
\label{sec:format}

\vspace{-0.1cm}The optic nerve in the human visual system contains axons that pass signal information from eyes to the brain. High intracranial pressure can cause these axons to swell, leading to papilledema. The diagnosis of papilledema in children at early stages is crucial because pediatric papilledema can be a sign of life-threatening conditions such as brain tumors and meningitis~\cite{mccafferty2017diagnostic}.  Several automated algorithms have been proposed recently to detect papilledema using fundus images. Milea et al.~\cite{milea2020artificial} demonstrated that deep learning can differentiate healthy optic discs from papilledema cases using fundus images with a clinically acceptable accuracy. Saba et al.~\cite{saba2021automatic} explored the effect of different deep neural network architectures on the eventual classification accuracy. Vasseneix et al~\cite{vasseneix2021accuracy} demonstrated that the diagnosis accuracy of AI-based methods is comparable to that of professional ophthalmologists~\cite{vasseneix2021accuracy}. However, these studies do not account for the critical confound arising due to the presence of pseduopapilledema in the input. Moreover, in the cases that do~\cite{ahn2019accuracy}, they primarily focus on adult subjects, where the differentiation between papilledema and pseudopapilledema is typically easier than in children. Indeed, anatomic differences in younger patients render pseudopapilledema more similar in appearance to papilledema~\cite{chang2017accuracy}.

Since pseudopapilledema is not a sign of a serious neurologic disorder \cite{chang2016optic}, distinguishing pseudopapilledema from papilledema at an early stage can help to circumvent the expensive and sometimes invasive clinical testing required for patients with papilledema. Although the ability to distinguish between the two is an invaluable need, works on this direction in human-centered AI-based medicine are limited. Despite demonstrating plausibility of using AI for this task, a few recent AI-based works use the standard black-box training regime of deep learning~\cite{ahn2019accuracy,milea2020artificial,liu2021detection} which makes their clinical applicability limited. Consequently, there is an unmet need to design explainable algorithms for this problem to screen patients who may not be examined immediately by ophthalmologists and also to assist clinicians in challenging diagnostic cases.

\begin{figure*}
    \centering
    \includegraphics[scale=0.49]{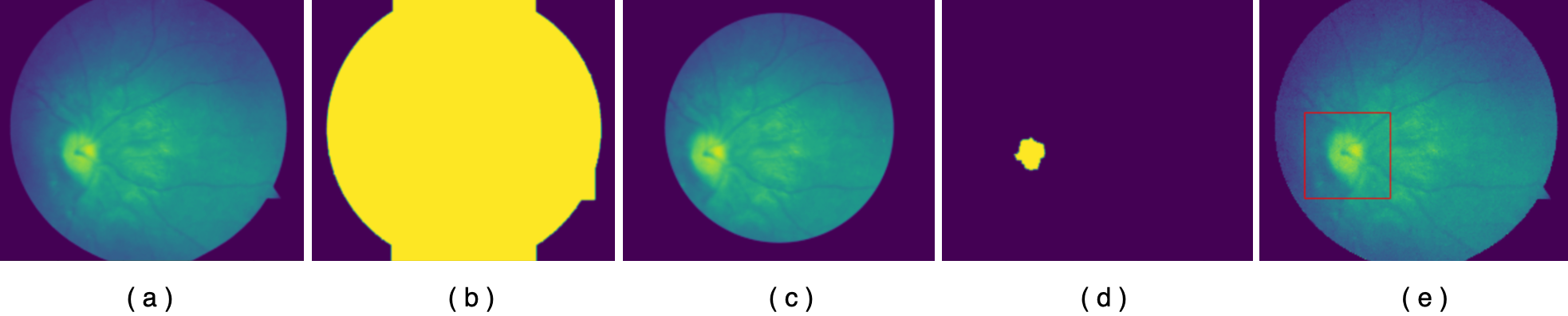}\vspace{-0.35cm}
    \caption{Unsupervised optic disc detection: a) Red channel of a sample fundus image after equalization. b) Binary mask generated after low thresholding and morphological operations. c) Result of fitting a circle and cropping at a smaller radius. d) Binary mask generated after high-value thresholding and morphological operations. e) Fitted bounding box to be cropped as the final optic disc proposal.}
    \vspace{-0.2cm}
    \label{fig:disc}
\end{figure*}

\section{Methodology}
\label{sec:method}

Our goal is to incorporate human-interpretable  physiological indicators into the deep learning framework by using appropriate image transformations. In Section~3.1 we introduce the selected indicators based on the clinical practice of distinguishing between papilledema and pseudopapilledema.
%These biomarkers are selected because clinicians use them to distinguish between papilledema and pseudopapilledema.
In Sections~3.2 and 3.3 we describe our algorithm to extract regions of clinical interest and the image transformations we use to emphasize the presence of the considered biomarkers in fundus images. Last, in Section~3.4 we describe the proposed deep learning architecture for papilledema detection.
%Using these transformations helps us to develop an algorithm with a human-centered and interpretable decision process.

\subsection{Papilledema Indicators}

%While clinicians have made substantial progress in detecting papilledema from fundus images against healthy controls, the clinical differentiation of papilledema from pseudopapilledema is more challenging. 
To overcome the challenge of clinical differentiation of papilledema from pseudopapilledema, Carta et al.~\cite{carta2011} conducted a machine learning study in adults and children with various causes of pseudopapilledema and optic disc edema (including papilledema) in which they came up with four robust signs on fundus examination that are specific to presence of papilledema:
\begin{itemize}
    \item \textbf{Swelling} (Fig.~1a): blurring of vessels at optic disc contour.
    \vspace{-0.1cm}
    \item \textbf{Hemorrhages} (Fig.~1b): peripapillary hemorrhages.
    \vspace{-0.1cm}
    \item \textbf{Elevation} (Fig.~1c): anterior extension of the optic nerve head on retinal surface.
    \vspace{-0.1cm}
    \item \textbf{Congestion} (Fig.~1d): enlargement and tortuosity of the arcuate and peripapillary venous vessels.
\end{itemize}
These clinical indicators can be located on, or at the perimeter of the optic disc, hence the first step of our algorithm is to detect and extract the optic disc as the region of interest.

\subsection{Optic disc Detection}

%In order to develop a machine learning architecture that extracts these human-centered and explainable indicators from fundus images, we must isolate the region of interest which is centered at the optic disc. 
Optic disc detection, localization and segmentation have been extensively investigated~\cite{sinthanayothin1999automated,akram2010,bajwa2019two}. Most works fine-tune deep learning models on binary masks of optic discs using a manually-annotated dataset \cite{milea2020artificial}. Given that we have access only to a small number of fundus images  without such fine-grained annotations in our dataset, we develop an unsupervised optic disc detection algorithm based on intensity thresholding and morphological operations.

Optic disc is typically one of the brightest regions in fundus images. This stems from concentration of blood vessels in that region, which mainly enhances intensity in the red and green channels of the RGB image. However, while intensity thresholding is a well-followed heuristic for optic disc detection \cite{dash2018unsupervised}, there could be other bright spots in the image, e.g., due to some disease or imperfect image capturing conditions, that can distort the detection outcome \cite{bajwa2019two}. Also, the green channel is most sensitive to outer vascular structures. Hence we isolate the red channel of each image and apply thresholding at 20\% of the mean intensity value so as to extract the greater region of the retina. We then apply sequential morphological opening and closing operations to shape the retina binary mask (Fig.~\ref{fig:disc}b). Afterwards, we fit a circle around that region and crop the encircled disc at a 20\% smaller radius. By doing so, we eliminate marginal noisy illumination. For the majority of cases, this is already a consistent isolation of the optic disc's greater region, as depicted in Fig.~\ref{fig:disc}c.

In order to improve detection, we further apply mean filtering to alleviate the effect of vascular structure and then threshold to mask the optic disc. The threshold was estimated empirically at 99\% of the maximum intensity value, so as to isolate only the brightest components. Again, we apply morphological filtering to form a cohesive disc region (Fig.~\ref{fig:disc}d) and fit an enlarged bounding box (Fig.~\ref{fig:disc}e) to account for possible misalignment. In addition to producing good detection results, our unsupervised approach also provides a way to scale up the task of optic disc detection. Specifically, we verify the area and eccentricity of the proposed disc mask (Fig.~\ref{fig:disc}d) in order to avoid false artefact detection. We thereby discard all components with no eccentricity and select the one with the highest area out of the remaining. If the condition is not met, the cropped image from the first step (Fig.~\ref{fig:disc}c) is returned as a better estimate. \vspace{-0.2cm}

\subsection{Augmentation Procedure}

After isolating the region of interest (Fig.~\ref{fig:features}), we use suitable image transformations to unveil papilledema indicators. We subsequently use these transformations in our model to contrast different input views of an image. The proposed scheme focuses on properties of separate color channels in fundus imaging. While the red channel could usually isolate the optic disc quite well, as described in Section 3.2, it can also depict hemorrhages or hyperemia in the greater optic disc region. These are also small-scale retinal blood structures that become brighter in the red channel. On the other hand, the green channel is mostly associated with perceived contrast and identification of vessels \cite{fraz_blood_2012}. Hence, it would be informative in detecting congestion and swelling. Finally, schematically identifiable features like elevation or hemorraghes can be located in any of the transforms or the original image under additive contrast.

Based on the these characteristics, we create three input views from each fundus image, to be trained together through a tri-branch model: the cropped image of the optic disc and the cropped images with adjusted contrast on the red and green channels. The proposed model is depicted in Fig.~\ref{fig:arch}. Each image is fed into a different pretrained branch to detect the respective features. We apply the transformations every time the sample is called during training. Each time, the contrast factor is a uniformly sampled value between 1.5 and 1.8. We specified the range of the contrast factor through cross-validation, aiming to produce substantially diverging but not distorted visual outcomes. During test time, images are augmented using the minimum factor value of 1.5.

\subsection{Architecture and Training}

\begin{figure}
    \centering
    \includegraphics[scale=0.33]{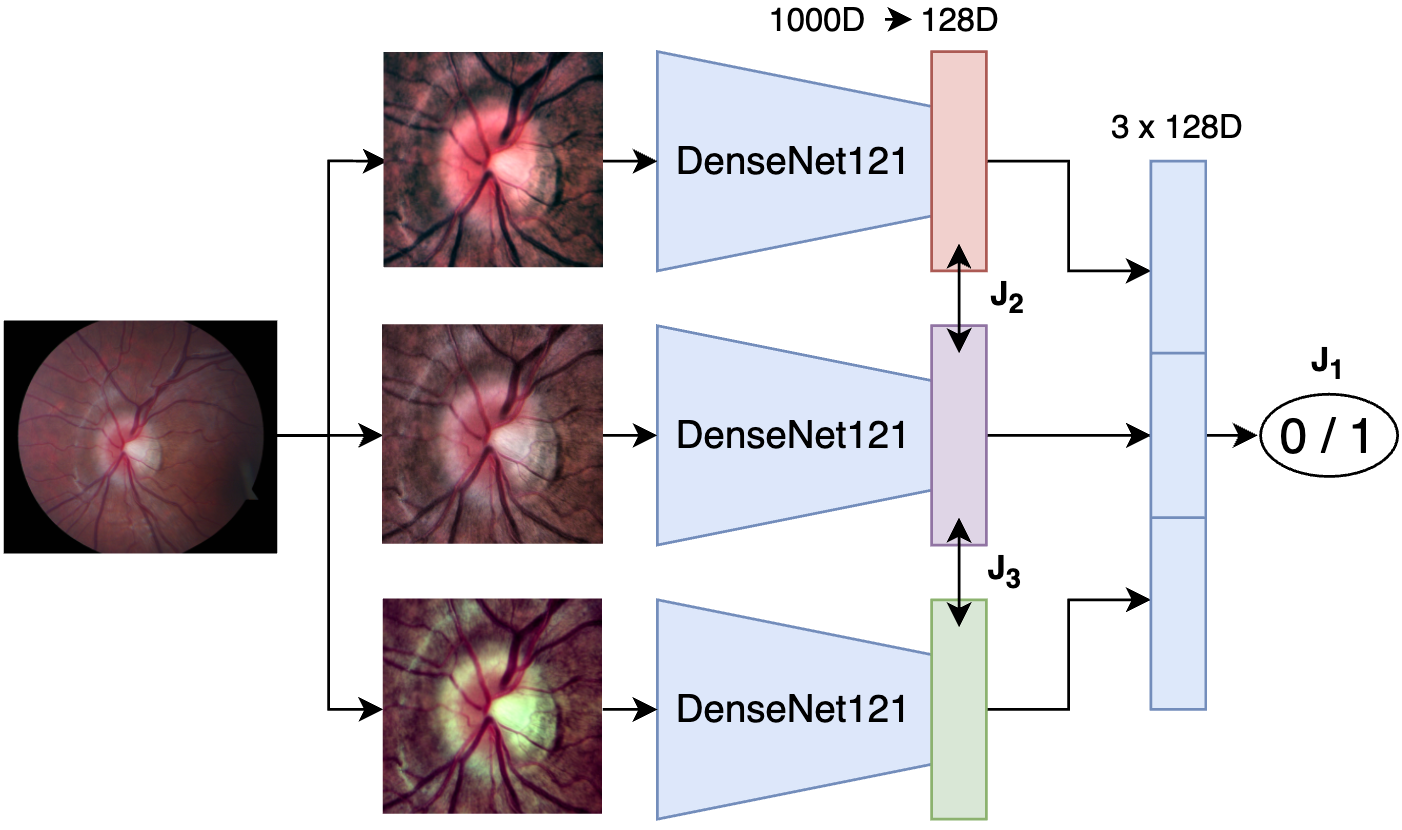}
    \vspace{-0.1cm}
    \caption{Proposed model tri-branch  architecture.}
    \vspace{-0.3cm}
    \label{fig:arch}
\end{figure}

We use a Densenet121 \cite{huang2017densely} for feature extraction in each branch of the proposed model (Fig.~\ref{fig:arch}) as it has been proven efficient in identifying papilledema in one of the most comprehensive AI studies~\cite{milea2020artificial} and is a prominent choice for medical tasks \cite{xiang2019towards, nguyen2021vindr}. Each branch has been pretrained on ImageNet and is fine-tuned on our small-scale fundus image dataset. The output feature vectors of each branch are projected to 128D embeddings, which are then concatenated to create an eventual feature vector for classification.

A combination of supervised and unsupervised loss functions is used to efficiently train the model for leveraging the augmented views of the input. Cross-entropy loss $\mathcal{J}_1(y,y^*)$ is used for label supervision, where $y \in \{0,1\}$ denotes the presence of pseudopapilledema or papilledema, and $y^*$ is the model output prediction. Two triplet losses $\mathcal{J}_2(v, v_{\text{red}}), \mathcal{J}_3(v, v_{\text{green}})$ contrast the 128D embeddings $v_{\text{red}}, v_{\text{green}}$ of the augmented views to that of the original image $v$. Positive pairs are created between the embedding of an original image and those of its two computed augmentations. Negative pairs are created by random assignments of augmented views to original images within a batch. Triplet losses then attempt to minimize the anchor's cosine distance to the positive example, while maximizing its distance to the negative one. We train first for 10 epochs using $\mathcal{J}_{\leq10}$ and then, until a total of 50, using $\mathcal{J}_{>10}$, where:
\vspace{-0.1cm}
\begin{equation}
    \mathcal{J}_{\leq10} = \lambda_2\,\mathcal{J}_2 + \lambda_3\,\mathcal{J}_3
\end{equation} \vspace{-0.5cm}
\begin{equation}
    \mathcal{J}_{>10} = \lambda_1\,\mathcal{J}_1 + \lambda_2\,\mathcal{J}_2 + \lambda_3\,\mathcal{J}_3
\end{equation}
In (1) we used a balanced combination of triplet losses, $\lambda_2 = \lambda_3$, whereas in (2) we emphasize supervised training with $\lambda_1 \gg \lambda_2, \lambda_3$. Each model instance is trained using Early-Stopping with a patience of 5 epochs of non-decreasing validation loss. AdamW~\cite{loshchilov2017decoupled} optimizer with a static learning rate of $1\text{e-4}$ is used. To avoid overfitting, the DenseNet backbones are kept frozen and we solely fine-tune their output embedding layers until convergence. \vspace{-0.1cm}

\section{Experimental Setup}
\label{sec:setup}

\vspace{-0.2cm}For the purposes of this study we collected a dataset of 331 pediatric fundus images that we obtain clinically from 105 subjects at Children's Hospital Los Angeles and University of California, Los Angeles between 2011 and 2021. All subjects were under 18 years of age at the time of diagnosis, specifically between 1 and 17 years old, with a median of eleven. 67 were females and the rest of them (36.2\%) males. All charts were reviewed to confirm that the children whose fundus photos were used met strict diagnostic criteria for papilledema or pseudopapilledema. Children diagnosed with papilledema were required to have neuroimaging evidence of an intracranial lesion (such as brain tumor) causing increased intracranial pressure, or a lumbar puncture with opening pressure greater than 28cm H2O. Children with pseudopapilledema were required to either have 1) normal neuroimaging and lumbar puncture with opening pressure less than 28cm H2O or 2) six months of follow-up showing no change in optic nerve appearance. The images were uploaded with de-identified clinical data to the HIPAA-compliant Research Electronic Data Capture (REDCap) database at USC. 149 out of 331 images were annotated as having papilledema, accounting for 44 subjects in total, and the rest (58\%) as having pseudopapilledema.

We use the whole dataset of 331 fundus images to train and validate our proposed model in multiple sessions. We first distinguish between the samples of different subjects to avoid information leakage from within-subject correlations. The resulting stratified splits contain approximately 75\% train and 25\% test sample images, chosen uniformly at random. We perform a 10-fold repeated cross-validation over 5 sessions in order to derive an average performance metric for our model. The accuracy of classifying papilledema against pseudopapilledema was evaluated by calculating accuracy scores and area under the receiver-operating-characteristic curve (AUC), in consistency with most studies in the field \cite{milea2020artificial,vasseneix2021accuracy, liu2021detection}. \vspace{-0.5cm}

\section{Results}
\label{sec:results}

\vspace{-0.2cm}The performance results of the proposed methods are shown in Table~\ref{table:1}. We report mean score and standard deviation among 5 repeats of 10-fold cross validation. The baseline approach includes a single pretrained DenseNet model, trained on the original dataset using a standard cross-entropy loss. We observe that even this simplistic approach can distingusih papilledema against pseudopapilledema at above 72\% on average, more than 3\% higher than the reported accuracy of human experts on our data and 6\% higher than human accuracy on external data~\cite{chang2017accuracy}. After performing optic disc detection and constrain the system in the region that the prominent papilledema indicators concentrate, the accuracy increases by 6\% to reach 78.3\% AUC and 78.1\% on accuracy. Furthermore, models' predictions are now less variable and show a reduced deviation from the mean score. By further applying our augmentation strategy during training, we reach 80\% in accuracy, 8\% above our baseline. It is noteworthy that the augmentation method applied to the original image set provides a rather slight improvement of 1\%.

\begin{table}
\centering
\begin{tabular}{l|c|c|c}
\label{table:1}
Method           & Crop & AUC & Accuracy \\ \hline
DenseNet121      & & 0.723 $\pm$ 0.009 & 0.722 $\pm$ 0.010 \\
DenseNet121      & \checkmark & 0.783 $\pm$ 0.014          & 0.781 $\pm$ 0.010                         \\
3xDenseNet121 & & 0.734 $\pm$ 0.007 & 0.735 $\pm$ 0.006 \\
3xDenseNet121 & \checkmark      & \textbf{0.798 $\pm$ 0.007}      & \textbf{0.801 $\pm$ 0.008}                        \\ \hline
Human Preds   &           &  N/A   & 0.69
\end{tabular}
\vspace{-0.2cm}
\caption{Performance of proposed algorithms at 10-fold cross-validation. Each experiment was repeated 5 times. Mean values and standard deviation of evaluation metrics are reported. Accuracy of human predictions on our data is reported for reference.} \vspace{-0.1cm}
\end{table}

\begin{table}
\centering
\begin{tabular}{l|c|c}
Method                       & AUC & Accuracy \\ \hline
Proposed                     &\textbf{0.798 $\pm$ 0.007}      & \textbf{0.801 $\pm$ 0.008}\\
Proposed w/ $\mathcal{J}_{\leq10}$ & 0.793 $\pm$ 0.010   &  0.795 $\pm$ 0.009    \\ \hline
DenseNet121 (2X : G)         & 0.781 $\pm$ 0.010    & 0.782 $\pm$ 0.009         \\
DenseNet121 (2X : R)        & 0.787 $\pm$ 0.009     & 0.785 $\pm$ 0.007         \\
DenseNet121 (2X : RG)         & 0.789 $\pm$ 0.008    &    0.788 $\pm$ 0.008
\label{table:2}
\end{tabular}
\vspace{-0.2cm}
\caption{Ablative experiments on branches of the proposed architecture and on the selected objective  at 10-fold cross-validation. Each cross-validation experiment was repeated 5 times. Mean values and standard deviation of evaluation metrics are reported.} \vspace{-0.5cm}
\end{table}

We report ablation experiments in Table~2 to further articulate the effectiveness of the proposed architecture and the training procedure, i.e., necessity of all model branches and all optimization terms for optimal performance. We first trained the architecture by omitting the use of contrastive losses on the augmented views. The averaged outcome does not point to a significant difference compared to the proposed approach. However, we observe that the training strategy we followed alleviates severe representational differences between the augmented views, hence it prevents sessions of severely deviating performance that we otherwise encounter. This is reported through the reduced deviation from the average scores. We also trained the proposed architecture (Fig.~\ref{fig:arch}) by fusing only 2 out of the 3 channels. We observe marginal non-significant improvements compared to the single-branch model, thus indicating the importance of incorporating all branches for effective predictions. This observation also implies that training with the single red or green augmented channel would not provide any further significant insights. \vspace{-0.2cm}

\section{Discussion}
\label{sec:print}

\vspace{-0.2cm}By comparing our results for automatic detection of papilledema versus pseudopapilledema against clinical predictions that were overseen by the authors, we observe that image processing systems based on deep learning are capable of at least matching human experts' capacity in clinical diagnosis.
%
%Accurate screening from fundus photography would be a valuable tool for clinicians, as it further advances this field to enable faster and timely interventions. Patients could also avoid stressful screening periods and the need for more expensive, specialized examinations, that are typically required (Section~\ref{sec:setup}) to rule out pseudopapilledema cases.
%
Our proposed methodology aims at
%matching human clinical experience with image processing and machine learning in order to enhance the recognition capacity and
providing clinicians with an explainable and trustworthy screening assistant. To this end, our focus on papilledema signs on the optic disc region as well as the utilized image transforms indeed show significant improvement from the baselines, both regarding the average performance, the convergence consistency, as well as the training and inference time. Models trained on the smaller cropped images were at least twice as fast as our baselines.

\begin{figure}
    \centering
    \includegraphics[scale=0.355]{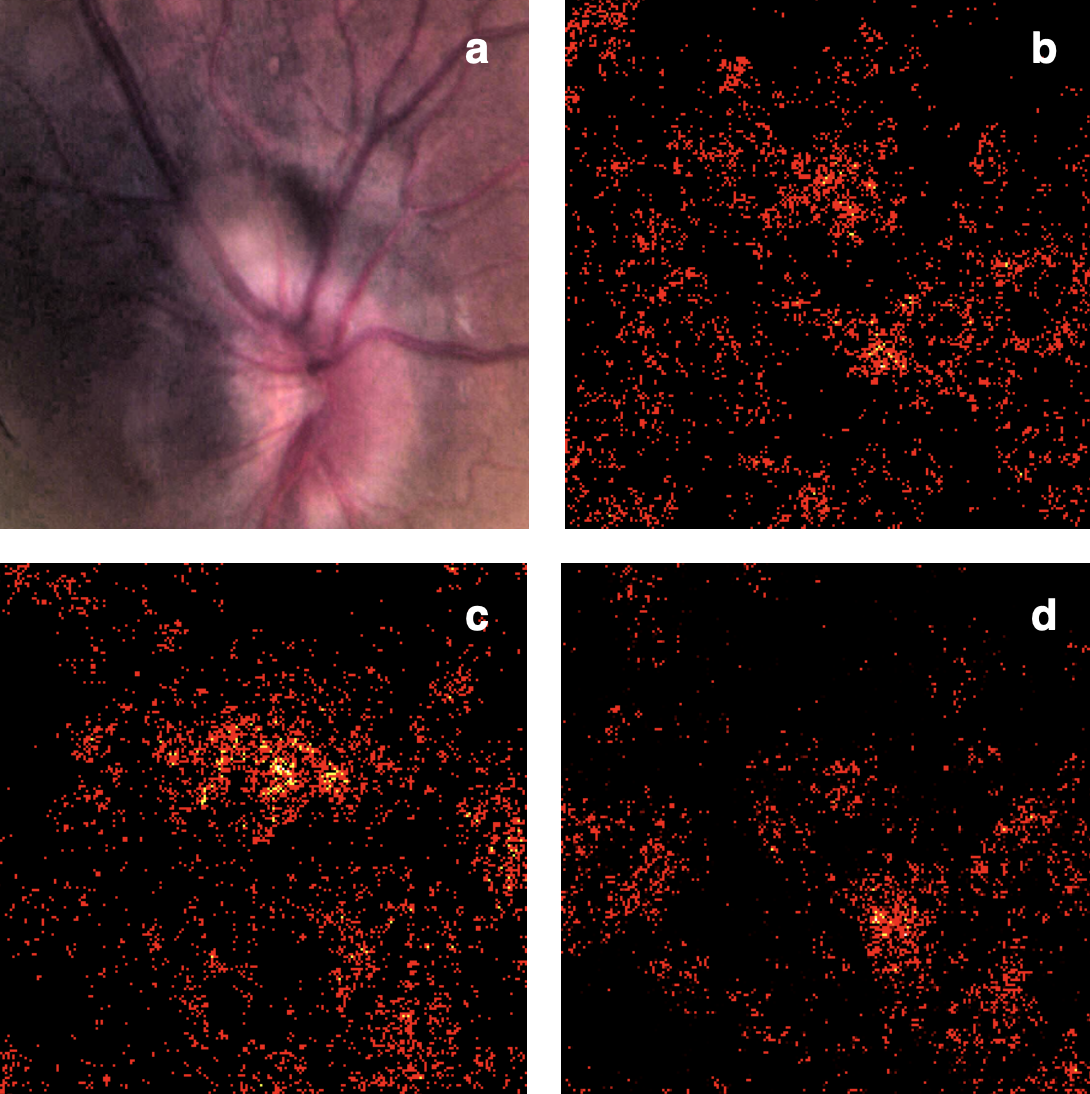}
    \vspace{-0.2cm}
    \caption{Salient activations of a test cropped fundus image (a). We depict the saliency map from the original input (b), the augmented red channel (c) and the augmented green channel (d).}
    \vspace{-0.5cm}
    \label{fig:head}
\end{figure}

We further extract sample saliency maps from random training sessions to illustrate the explainability of our approach. In Fig.~\ref{fig:head} we depict a sample fundus image along with the activation of each of the network branches. The saliency maps are indicative of present signs of papilledema in accordance with our analysis and were clinically verified. Specifically, the augmented-red channel activation highlights a superotemporal cotton wool spot, which is a sign of papilledema, similar to retinal structures that are shown brighter in the red channel. The augmented-green channel map corresponds to swelling of the retinal nerve fiber layer, best seen on the nasal aspect of the optic nerve. These activations are stronger in the respective branches than on the non-augmented branch, shown in Fig.~\ref{fig:head}b.

%\begin{figure}
%    \centering
%    \includegraphics[scale=0.34]{img/head2.png}
%    \vspace{-0.3cm}
%    \caption{Papilledema vs pseudopapilledema classification for the proposed algorithms. Green dashed line represents the human experts' capability on the task. We demonstrate that a combination of clinically-oriented explainable features and augmentation methods can enhance automatic detection.}
%    \vspace{-0.4cm}
%    \label{fig:head}
%\end{figure}

%Self-supervised learning has been expanding in multiple areas of deep learning research and has shown advances in medical imaging as well. Azizi et al. \cite{azizi2021big} showed that downstream performance on medical imaging tasks can be greatly improved when neural networks are trained in multiple self-supervised sessions, first on large-scale general datasets and then on specialized medical data. The training algorithm we utilized in this study is essentially an adaptation of this notion to our small-scale task. Despite the slight improvement shown at the ablation study, this method could stabilize the learning process and we firmly believe it can scale efficiently in the presence of a larger number of data in our future work.

Our study has limitations as well. As mentioned above, the utilized dataset is still of small scale and, as a result, the proposed algorithms were implemented using repeated cross-validations, extensive regularization and largely frozen pretrained backbones, in order to avoid overfitting and false within-subject correlations. This poses questions regarding the generalization capability of the model, however we expect that the suggested augmentation scheme and multi-objective training will scale efficiently as new data populate our database \cite{azizi2021big}. Moreover, it is a reliable assisting model in cases of data scarcity. Another limitation concerns the absence of image metadata like biomarkers or optic disc masks that could help substitute handcrafted detection with an automatic machine learning process. Nevertheless, our proposed approach still shows robust performance against the baselines and offers proof of concept. \vspace{-0.15cm}

\section{Conclusion}
\label{sec:foot}
\vspace{-0.15cm}
In this study we presented a deep learning-based approach for the challenging task of detecting and differentiating pediatric papilledema from pseudopapilledema in fundus images. We constructed an interpretable system based on clinical papilledema indicators that could be used as a trustworthy screening assistant in the hands of experts. We proposed an unsupervised algorithm for optic disc detection and a contrastive training strategy to unveil important features through deep learning. The obtained results suggest that our system is capable of detecting true papilledema at an accuracy that is comparable to clinical standards. In the future we will expand our database and scale our framework to assess its robustness. We also intend to ground our study on clinical indicators using fine-grained annotations, that would allow more explicit evaluation.

%\section{Acknowledgements}
%\label{sec:copyright}

\vfill\pagebreak
{\small
\bibliographystyle{IEEEbib}
\bibliography{refs}
}

\end{document}